\documentclass[11pt]{article}
\usepackage{moriond,epsfig}

\def\Journal#1#2#3#4{{#1} {\bf #2}, #3 (#4)}

\newcommand{\ARAA}{ARA\&A}
\newcommand{\AaA}{A\&A}

\def\be{\begin{equation}}
\def\ee{\end{equation}}
\def\bea{\begin{eqnarray}}
\def\eea{\end{eqnarray}}

\begin{document}
\vspace*{4cm}
\title{NON-GAUSSIAN SIGNATURE INDUCED BY THE SZ EFFECT OF GALAXY CLUSTERS}

\author{ N. AGHANIM \& O. FORNI }

\address{Institut d'Astrophysique Spatiale, B\^at. 121, Universit\'e Paris 
Sud,\\
91405 Orsay Cedex, France}

\maketitle\abstracts{We apply statistical tests, based on the study of the
coefficients in a wavelet decomposition, to a cosmological signal: the 
Cosmic Microwave 
Background (CMB) anisotropies. The latter represent the superposition of 
primary anisotropy imprints of the initial density perturbations and 
secondary anisotropies due to photon interactions after recombination. In an 
inflationary scenario with Gaussian distributed 
fluctuations, we study the statistical signature of the
secondary effects. More specifically, we investigate the dominant effects 
arising from the Sunyaev-Zel'dovich effect of galaxy clusters.
Our study predicts the non-Gaussian signature of these secondary 
anisotropies and its detectability in the context of the future CMB 
satellite Planck Surveyor.}

\section{Introduction}
One of the major goals  of cosmology is to understand the origin of the 
initial density perturbations. Do they come from inflation\cite{guth81} or 
from topological defects\cite{vilenkin85}? One way to answer this question is 
to make a statistical analysis of the CMB anisotropies. In fact, inflation 
predicts a Gaussian distribution of the primary anisotropies whereas the 
topological defects predict a non-Gaussian distribution. However, not only 
topological defects produce 
non-Gaussian signals. There are other astrophysical sources of non-gaussianity 
such as gravitational lensing (cf. F. Bernardeau's contribution in this 
proceedings) and the secondary anisotropies such as the Sunyaev-Zel'dovich 
effect\cite{sunyaev80} of galaxy clusters. \\
In this context, we have 
developed a method for the statistical analysis based on the wavelet 
decomposition of a signal. The method has been tested on an arbitrary set of 
Gaussian and non-Gaussian maps and then applied to CMB simulated maps to 
investigate the effect of the SZ contribution on the statistical signature of
the anisotropies. 
\section{Detection method}
Our method is based on a wavelet analysis of an image. That is, rather than 
analysing the signal in the direct space, we analyse the associated 
coefficients in the wavelet 
decomposition. The wavelet analysis can be viewed as a convolution of the 
signal by band pass filters: a scaling function similar to a low pas filter 
and a wavelet function similar to a high pass filter. The low pass filter will 
give an image with a lower resolution and the high pass will give the 
associated details. We have used the multi-scale wavelet 
analysis\cite{mallat89} to decompose 
an input signal into a series of successive low resolution images and their 
associated detail images. At each level of the decomposition, the reference 
image has a resolution reduced by a factor of two and only this reference low 
pass image is decomposed at each level. This is the principle of the dyadic 
decomposition. We
have chosen an anti-symmetric wavelet\cite{villasenor95} function similar to a 
first derivative 
operator because we expect that the non-Gaussian signatures arise from sharp 
gradients in the signal. Such a function is particularly sensitive to 
gradients and will detect them easily. The high pass filter applied to both
 directions of the image gives at each level three detail images corresponding 
to vertical, horizontal and diagonal details. The features of interest, 
particularly the statistical signatures, are studied at each scale (or level) 
for each type of details.
We have qualified and tested our detection method on a set of Gaussian and 
non-Gaussian maps (100 of each) having the same power spectrum. This condition 
allows us to attribute the differences that are detected between the 
Gaussian and non-Gaussian maps to nothing but their statistical nature.\par
Our detection method\cite{forni99} is based on the three following major 
steps. First, we perform 
the wavelet decomposition of both 100 Gaussian maps and 100 non-Gaussian maps. 
We thus obtain at each decomposition level the wavelet coefficients 
associated with the vertical, horizontal and diagonal details. We also define 
the multi-scale gradient as the sum of the squared coefficients associated 
with the vertical and horizontal details. The non-Gaussian signature can be
detected through the third or fourth order moment of the distribution, 
respectively the skewness and the kurtosis. In the following, we focus only 
on the excess of kurtosis because our signals are very moderately skewed.
The second step of the method is to compute for 
each Gaussian and non-Gaussian map and at each level the excess of kurtosis of 
the wavelet coefficients associated with the corresponding details. We 
check that contrary to the Gaussian maps, the excess of kurtosis of the
non-Gaussian signal is always 
centred around a non-zero value. We also note that the dispersion 
around the mean excess of kurtosis increases at high decomposition levels even 
for the Gaussian maps, due to the lack of coefficients at these levels. 
Therefore, in the following, we restrict our study to the first three levels.
 \par
The last step of the analysis is to quantify the detectability of the 
non-Gaussian signature. This is done by comparing the probability distribution 
functions (PDF) of the different processes. We compute the median excess of 
kurtosis of the 100 non-Gaussian maps at each decomposition scale and 
estimate the probability that it belongs to the PDF of the Gaussian maps. A 
low probability favours of a non-Gaussian signal whereas a high probability 
indicates a Gaussian process. We can also perform a more global comparison of 
the PDFs through the Kolmogorov-Smirnov test\cite{press92}. It gives, with a 
very good accuracy, the probability for a distribution to be different from a 
Gaussian.
\section{Application to CMB}
We apply our method to simulated CMB data including primary and secondary 
anisotropies. Our goal is to estimate the statistical non-Gaussian 
contamination induced by the secondary anisotropies. Therefore, we use an 
inflation model that generates Gaussian distributed primary anisotropies to 
which we add the simulated\cite{aghanim98} contributions due to the 
Sunyaev-Zel'dovich (thermal and kinetic) effect of galaxy clusters. \par
We have simulated 100 statistical realisations of the resulting CMB maps and 
performed the multi-scale decomposition. Following our proposed method, we 
compute the excess of kurtosis at different scales for the coefficients 
associated with the diagonal, vertical and horizontal details (Table 1).
\\
 At the first three decomposition
scales, the excess of kurtosis is very large due to the SZ 
contribution. We also note,
that the computations using the diagonal details are more sensitive 
to non-gaussianity and thus more powerful in detecting it. In 
fact, the galaxy clusters exhibit very peaked profiles or even point-like 
behaviour. The diagonal details are very sensitive to symmetric profiles. We
find that the SZ effect a major source of non-gaussianity among the secondary 
effects.

\begin{table}
\label{tab:cbsz_derv}
\begin{center}
\begin{tabular}{|c|c|c|c|c|c|c|c|c|}
\hline
Scale & & $k_1$ & $\sigma_+$ & $\sigma_-$ & & $k_2$ & $\sigma_+$ & $\sigma_-$\\
\hline
I & $\partial/\partial x$ & 13.89 & 11.67 & 3.97 & & 11.98 & 7.66 & 3.74 \\
II &  \& & 2.90 & 1.29 & 0.48 & $\partial^2/\partial x\partial y$ & 6.88 & 7.00 & 1.73\\
III & $\partial/\partial y$ & 0.09 & 0.08 & 0.08 & & 0.36 & 0.17& 0.11 \\
\hline
\end{tabular}
\end{center}
\caption{The median excess of kurtosis, at four decomposition 
scales, computed over 100 realisations of the sum of CMB and
secondary anisotropies (inhomogeneous re-ionisation$^1$ and 
thermal and kinetic SZ 
effect). The signal includes the Gaussian noise expected for the Planck 
mission. $k_1$ is the median excess computed with the coefficients
associated with the vertical and horizontal gradients, and $k_2$ is given for 
the diagonal details. The $\sigma$ values are the boundaries
of the confidence interval for one statistical realisation.}
\end{table}

\section{Effects of the instrumental configurations}
We apply our statistical discriminators to test for non-gaussianity within the 
context of the representative instrumental configuration of the future  Planck 
Surveyor satellite for CMB observations. 

\begin{figure}
\epsfxsize=\columnwidth
\hbox{\epsffile{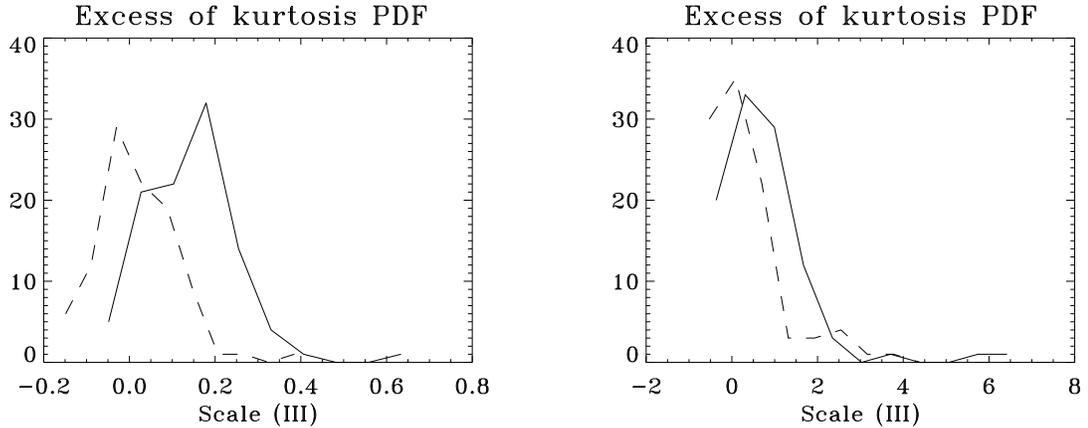}}
\caption{For the Planck-like configuration (CMB + SZ + inhomogeneous
 re-ionisation), probability distribution functions, as percentages, of the 
excess of 
kurtosis computed with the multi-scale gradient coefficients (right panels)
and with the coefficients of the diagonal details (left panels). The dashed line is
for the Gaussian test maps with same power spectrum as the non-Gaussian signal 
(solid line).}
\label{fig:pdfpl}
\end{figure}

We use the same astrophysical contributions as those described above (primary 
and SZ). The difference lies in the fact that the  maps are convolved 
with a 6 arcminute Gaussian beam. We also take into account the expected 
Gaussian noise of Planck ($(\delta T/T)_{rms}\sim2.\,10^{-6}$ per 1.5 
arcminute pixel). The convolution by a 6 arcminute beam suppresses the 
power at the corresponding scale (Scale I) and affects the second 
decomposition scale whereas the third is not significantly altered. At the 
third decomposition scale, we find for the multi-scale gradient 
$k=0.62^{+1.43}_{-0.60}$.
Whereas we find for the horizontal and vertical details, and for the diagonal
details respectively, 
$k=0.07^{+0.11}_{-0.08}$ and $0.16\pm0.10$.  In order to quantify the 
detectability of non-gaussianity
in the Planck-like configuration, we generate Gaussian distributed maps with 
same power spectrum as the 
studied signal. We plot (Fig. \ref{fig:pdfpl}) the PDF of the Gaussian (dashed 
line) and non-Gaussian (solid line) 
processes. We derive the probability that the median excess of kurtosis 
measured on the ``real sky'' belongs to
the Gaussian process. Using the multi-scale gradient, we find that the 
probability of detecting non-gaussianity is
71.9\% at the second decomposition scale. There is no 
significant detection elsewhere. Whereas using the coefficients of the
diagonal details, the probability of detecting a non-Gaussian signature at the 
third scale is 94.5\%. We apply the K-S test to the distribution of the excess
of kurtosis for the diagonal details and find a probability of 96.6\% of
detecting non-gaussianity. Since the comparison of the two distributions using 
the K-S test is very sensitive to departures from gaussianity. It thus gives 
better results on the detection of the non-Gaussian signature.
\section{Conclusions}

In the present study, we investigate the statistical signature induced by the 
SZ effect of galaxy clusters when the primary anisotropies result from an 
inflationary scenario and are thus Gaussian distributed. We use 
discriminators based on the statistical properties of the coefficients in a 
four level wavelet decomposition.
In our study, we find that the SZ effect of clusters generates a very large
non-Gaussian signature that dominates by far all other secondary anisotropies.
We apply our statistical tests to a Planck-like  
configuration in order to estimate the capabilities of the 
satellite to detect the non-Gaussian signature of the SZ effect.
In this case, we detect unambiguously the non-Gaussian signature at the third
decomposition scale ($\simeq 12$ arcminutes), the first 
($\simeq 3$ arcminutes) and second ($\simeq 6$ arcminutes) scales being 
affected by the beam convolution.

\end{document}